\documentclass[aps,prl,twocolumn,letterpaper,reprint,superscriptaddress]{revtex4}
\usepackage{amssymb}
\usepackage{amsmath}
\usepackage{graphicx}
\usepackage{hyperref}

\def\rd{R_{\mathrm{D}}}

\def\p0{\phi_0}
\def\a0{A_0}

\def\um{\mathrm{\mu m}}

\def\Tesla{\mathrm{T}}
\def\mt{\mathrm{mT}}
\def\mv{\mathrm{mV}}
\def\mk{\mathrm{mK}}

\def\fc{f_\mathrm{c}}
\def\fb{\nu_\mathrm{b}}

\def\invr{1/r}

\def\ecs{e^*}

\def\alith{A_\mathrm{lith}}

\def\vb{V_{\mathrm{B}}}
\def\dvt{\Delta V_{\mathrm{T}}}
\def\dvb{\Delta V_{\mathrm{B}}}
\def\dvlt{\Delta V_{\mathrm{LT}}}
\def\dvrt{\Delta V_{\mathrm{RT}}}
\def\vt{V_{\mathrm{T}}}
\def\vb{V_{\mathrm{B}}}
\def\vlt{V_{\mathrm{LT}}}
\def\vrt{V_{\mathrm{RT}}}
\def\vlb{V_{\mathrm{LB}}}
\def\vrb{V_{\mathrm{RB}}}
\def\dvg{\Delta V_{\mathrm{g}}}
\def\vg{V_{\mathrm{g}}}
\def\gg{\gamma_{\mathrm{g}}}
\def\bg{\beta_{\mathrm{g}}}
\def\eg{\eta_{\mathrm{g}}}

\def\ceff{C_\mathrm{eff}}

\def\rxy{R_{\mathrm{xy}}}

\def\Db{\Delta B}

\def\df{\delta f}

\def\f0{f_\mathrm{0}}
\def\nb{n_\mathrm{b}}
\def\nl{N_\mathrm{L}}
\def\nbg{N_\mathrm{BG}}
\def\dbg{n_\mathrm{BG}}
\def\nphi{N_\mathrm{\phi}}

\def\percm2{\mathrm{cm}^{-2}}
\def\perm2{\mathrm{m}^{-2}}

\begin{document}
\title{Fabry-Perot Interferometry with Fractional Charges}
\author{D.\ T.\ McClure}
\affiliation{Department of Physics, Harvard University, Cambridge, Massachusetts
02138, USA}

\author{W.\ Chang}
\affiliation{Department of Physics, Harvard University, Cambridge, Massachusetts
02138, USA}

\author{C.\ M.\ Marcus}
\affiliation{Department of Physics, Harvard University, Cambridge, Massachusetts
02138, USA}

\author{L.\ N.\ Pfeiffer}
\affiliation{Department of Electrical Engineering, Princeton University, Princeton, New Jersey 08544, USA}

\author{K.\ W.\ West}
\affiliation{Department of Electrical Engineering, Princeton University, Princeton, New Jersey 08544, USA}
\date{\today}

\begin{abstract}
Resistance oscillations in electronic Fabry-Perot interferometers near fractional quantum Hall (FQH) filling factors $1/3, 2/3, 4/3$ and $5/3$ in the constrictions are compared to corresponding oscillations near integer quantum Hall (IQH) filling factors in the constrictions, appearing in the same devices and at the same gate voltages. Two-dimensional plots of resistance versus gate voltage and magnetic field indicate that all oscillations are Coulomb dominated. Applying a Coulomb charging model yields an effective tunneling charge $e^{*}\approx e/3$ for all FQH constrictions and $e^{*}\approx e$ for IQH constrictions. Surprisingly, we find a common characteristic temperature for FQH oscillations and a different common characteristic temperature for IQH oscillations. 
\end{abstract}

\maketitle

Like their optical analogs, electronic Fabry-Perot interferometers allow quantum interference to be probed via tunable parameters that induce periodic transmission oscillations. By working with charged excitations in quantum Hall edge states, however, the electronic version also allows the interplay of coherence, interaction, and magnetic effects to be studied; notably, such devices could provide direct evidence for anyonic~\cite{chamon97} and non-Abelian~\cite{stern06,bonderson06,ilan08,stern09} statistics and potentially form the building blocks of topological quantum computers. In the integer quantum Hall (IQH) regime, where these devices have been studied for over two decades, recent experimental~\cite{godfrey07,caminoPRB07,yiming09,ofek10} and theoretical~\cite{bernd07,bert10} work has extended the results of earlier experiments~\cite{vanwees89, alphenaar92, mceuen92, taylor92, bird96} and highlighted the role of Coulomb interaction in producing the observed oscillations. More recently, oscillations consistent with Aharonov-Bohm (AB) interference of non-interacting electrons have also been observed~\cite{yiming09,ofek10}, and can be qualitatively distinguished from the Coulomb-dominated (CD) type using a 2D plot of resistance versus magnetic field and gate voltage.

In the fractional quantum Hall (FQH) regime, signatures of fractional charge~\cite{bernd07} as well as both Abelian~\cite{chamon97,bert10} and non-Abelian~\cite{stern06,stern09} statistics have been predicted in both the CD~\cite{stern06,bernd07,stern09,bert10} and AB~\cite{chamon97,stern06,stern09,bert10} regimes, but few experimental results have been published. With the two constrictions of a quantum-dot interferometer near filling factor $\fc=1/3$, Camino et~al.~\cite{caminoPRL07} observed oscillations consistent with tunneling of either $e/3$ quasi-particles in the CD regime or electrons in the AB regime~\cite{yiming09}; Ofek et~al.~\cite{ofek10} later reported a similar result with a 2D plot confirming the CD nature of the oscillations. Most recently, oscillations have been reported~\cite{willett10} near $\fc=7/3$ and $5/2$, though their unusually broad Fourier transforms suggest that they do not arise from particles encircling a well-defined area. These experiments~\cite{willett10}, as well as shot-noise measurements near FQH states in the first~\cite{chung03,bid09} and second~\cite{dolev10} Landau levels, suggest the possibility of tunneling mediated by quasi-particles with a larger charge than expected. Analysis of CD oscillations can reveal the charge of tunneling quasi-particles, but such measurements have not been reported for $\fc$ other than $1/3$, where experiments have consistently found the expected quasi-particle charge.

In this Letter, we report measurements of CD oscillations near the low-field edges of IQH and FQH plateaus at $\fc=r=1, 2, 3, 4$ and $\fc=r/3=1/3, 2/3, 4/3, 5/3$ in the constrictions. Observed dependence of gate-voltage periods on $r$ in each regime is well described by the charging model presented in Refs.~\cite{bernd07,bert10}. Using this model, we extract effective charges consistent with  $e^* \approx e/3$ for the fractional $\fc$ values, and  $e^* \approx e$ for the integer $\fc$ values. Magnetic field periods were found to be roughly proportional to $1/r$ in both the integer and fractional regimes, as expected from the charging model. The temperature dependence of oscillations was found to be exponential, as anticipated theoretically~\cite{bert10}, but with a surprising pattern: Two characteristic temperature scales were found, one for integers and one for fractions, across a range of $\fc$ values and  device sizes. These scales are not understood.

\begin{figure}[b!]
\center \label{figure1}
\includegraphics[width=3.25 in]{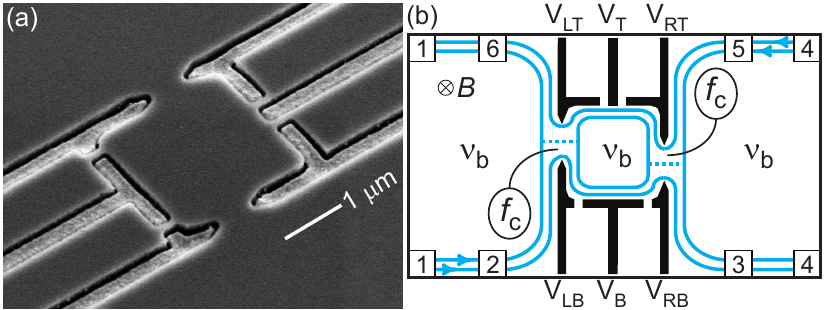}
\caption{\footnotesize{(a) Scanning electron micrograph of a $4~\um^2$ interferometer similar to those measured. (b) Gate layout of the $4~\um^2$ device with schematic diagram of edge state paths, filling factors, and ohmic contacts. A current bias applied between contacts 1 and 4 is used to measure the diagonal resistance $\rd$ between contacts 2 and 5, as well as the bulk Hall resistance $\rxy$ between contacts 3 and 5. This picture assumes that only one quantum Hall edge is partially transmitted by the interferometer, while others are fully transmitted or reflected; for clarity, only one fully-transmitted edge and no fully-reflected edges are shown.}}
\end{figure}

Interferometers were fabricated using e-beam lithography on GaAs/AlGaAs heterostructures with a two-dimensional electron gas (2DEG) of density $\nb=1.7\times10^{11}~\percm2$ and mobility $\mu = 2\times10^{7}~\mathrm{cm}^2/\mathrm{Vs}$ in a 40-nm quantum well centered $290~\mathrm{nm}$ below the surface. A BCl$_3$ reactive ion etch was used to form 150~nm deep trenches~\footnote{We believe trenches improve device performance in two ways: the steep confining potential allows the formation of sub-micron constrictions with $\fc\approx\fb$, and the elimination of the donor layer between the gates and the 2DEG enhances stability.} into which Ti/Au gates were deposited in the same lithographic step [Fig.~1(a)]. Measurements on two devices are reported, one with lithographic area $\alith=4~\um^2$ and 750 nm constrictions [identical to the device in Fig.~1(a), and shown schematically in Fig.~1(b)] and the other with $\alith=2~\um^2$, 600 nm constrictions, and a single gate $\vb$ in place of gates $\vlb, \vb, \vrb$. Devices were cooled in a dilution refrigerator with base temperature $\lesssim 10~\mathrm{mK}$~\footnote{Evidence that the 2DEG also reaches this temperature is presented in Fig.~5 and the associated discussion.}. The diagonal resistance $\rd$ across the interferometer and the bulk Hall resistance $\rxy$ were measured simultaneously using a lock-in with ac current bias $I = 0.25~\mathrm{nA}$ [Fig.~1(b)].

\begin{figure*}[t]
\center \label{figure2}
\includegraphics[width=7 in]{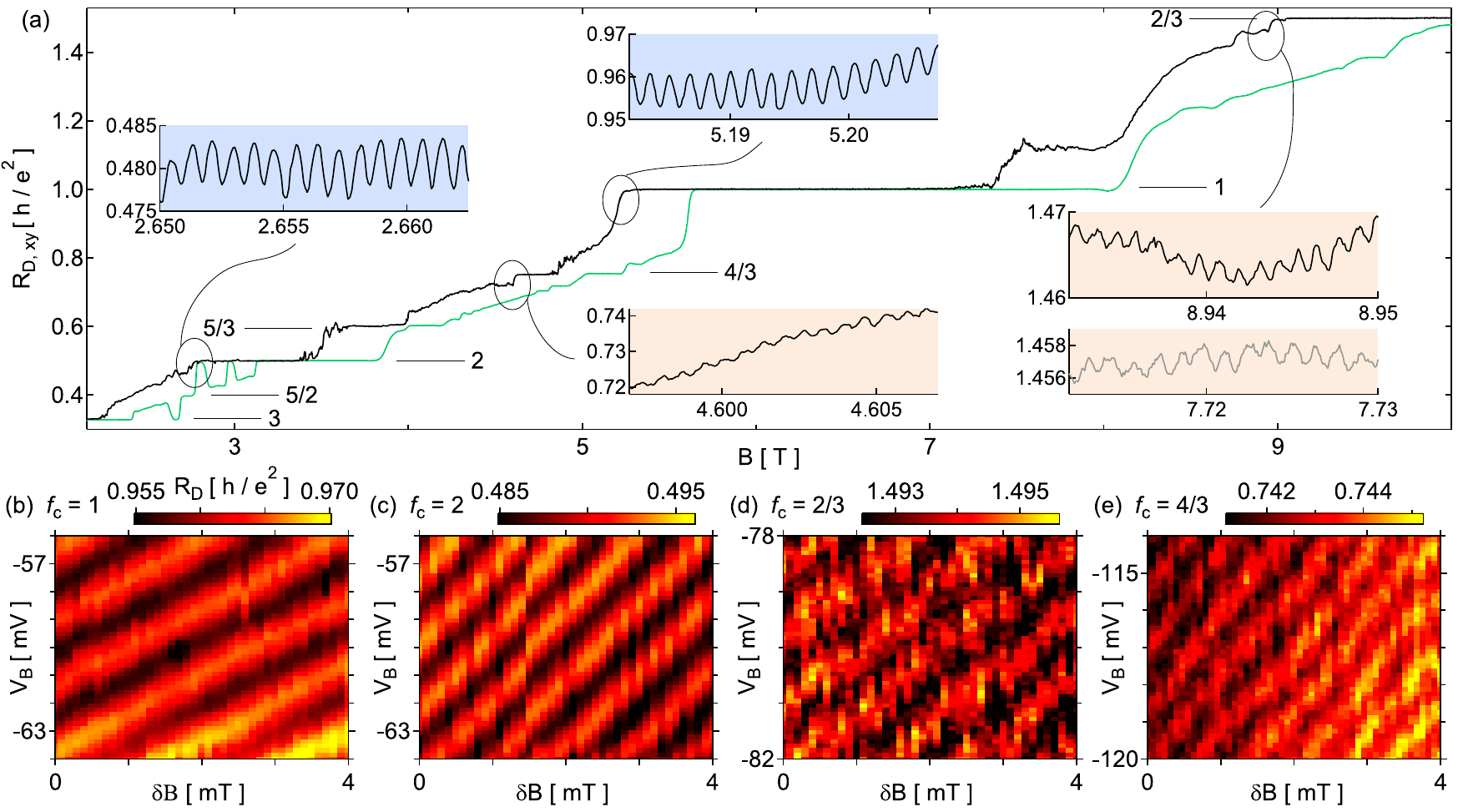}
\caption{\footnotesize{(a) Diagonal resistance $\rd$ (black) and bulk Hall resistance $\rxy$ (green) as a function of perpendicular magnetic field $B$, with $\vt$ and $\vb$ near $-100~\mv$ and all other gate voltages near $-200~\mv$. Numbered horizontal lines indicate notable filling factors. Insets: detail views of the $\rd$ trace, revealing oscillations at $\fc=1~\mathrm{(top)}, 2~\mathrm{(left)}, 2/3~\mathrm{(right)},$ and $4/3~\mathrm{(bottom)}$. For the lower panel in the $\fc=2/3$ inset, constriction gate voltages are near $-500~\mv$. All features are independent of the field sweep rate (typically $\sim20~\mt/\mathrm{min}$) and direction. Here and throughout, integer (fractional) $\fc$ are represented by blue (orange). (b-e) Plots of $\rd$ in the $B-\vb$ plane, with gate voltages comparable to those in (a). Here, $B=B_0 + \delta B$, with $B_0 = 5.200~\Tesla, 2.670~\Tesla, 8.831~\Tesla$, and $4.684~\Tesla$, respectively.}}
\end{figure*}

\begin{figure}
\center \label{figure3}
\includegraphics[width=3.25 in]{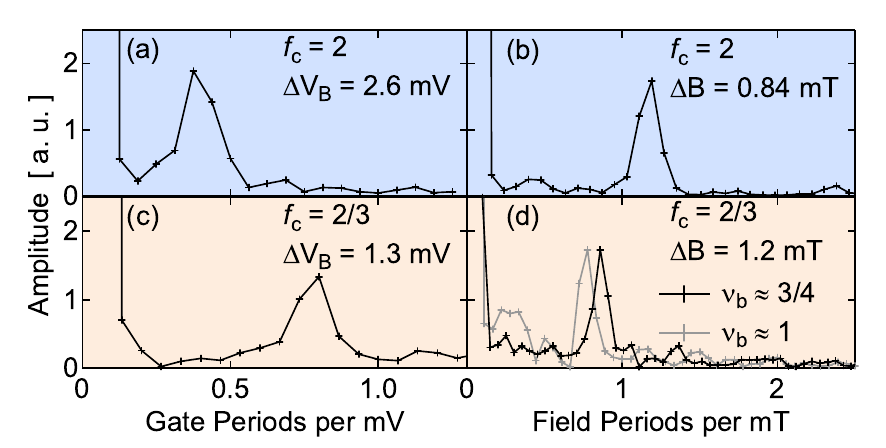}
\caption{\footnotesize{Sample FFT's of oscillations with respect to $\vb$ and $B$, for $\fc=2~\mathrm{and}~2/3$. The raw data for (b) and (d) are shown in the corresponding Fig.~2(a) insets, while the raw data for (a) and (c) are vertical cuts from 2D plots like those in Figs.~2(c,d) but with a larger gate-voltage range for greater frequency resolution.}}
\end{figure}

Figure~2(a) shows $\rxy$ and $\rd$ of the $4~\um^2$ device as a function of perpendicular magnetic field $B$, covering a range of filling factors from $2/3$ to $3$ in both constrictions and bulk. Voltages of $\sim -200~\mathrm{mV}$ on gates $\vlt,\vrt,\vlb$, and $\vrb$ reduced electron density in both constrictions by $\sim10\%$ compared to the bulk, while preserving several FQH plateaus. Oscillations in $\rd$ (insets in Fig.~2) were observed at the low-field edges of several IQH and FQH plateaus, where presumably the interfering edge is strongly backscattered and transport occurs via weak forward tunneling, as illustrated in Fig.~1(b). Gate-voltage adjustments allowed oscillations at each $\fc$ to be seen over a range of bulk filling factors $\fb$, as illustrated in the $\fc=2/3$ inset. Even as $\fb$ is tuned continuously through a range of compressible and incompressible states, the magnetic-field and gate-voltage periods remain nearly constant. The slightly larger $\Db$ in the lower panel of the inset [more clearly visible as a smaller frequency in Fig.~3(d)] can be understood as resulting from a slight decrease in device area at more negative gate voltages. Two-dimensional sweeps of magnetic field and gate voltage [Figs.~2(b-e)] show a positive slope of constant phase, indicating Coulomb-dominated oscillations~\cite{yiming09}.

Field and gate periods were extracted from fast Fourier transforms (FFT's) [Fig.~3], which all show a sharp peak at a single frequency. A gaussian fit to the peak gives the center frequency $\f0$ and full width at half-maximum (FWHM) $\df$, with periods $\Db$ or $\dvg$ given by $1/\f0$. For FFT's comprising $N_\mathrm{osc}$ oscillations, we find $\df\sim 1/N_\mathrm{osc}$, indicating that the peak width is limited by the finite data range.

Similar behavior to the data in Figs.~2 and 3 was observed in the $2~\um^2$ device, and at $\fc=1/3, 5/3, 3~\mathrm{and}~4$. The remaining figures present three data sets, with the same gate voltages used at all integer and fractional $\fc$ within each data set. Gate periods in Fig.~4(a) are normalized by their values at $\fc=1$, allowing comparison of periods from all four gates common to both devices. (Raw periods differ by up to an order of magnitude, as expected from the differences in gate sizes.) A steady increase in $\dvg$ with $\fc$ appears in the FQH regime, with a similar but weaker trend in the IQH regime. Values of $\Db$ are plotted in Fig.~4(b) as a function of $\invr$, where $\fc=r$ in the IQH regime and $\fc=r/3$ in the FQH regime. Separate fit lines constrained to cross the origin, representing $\Db\propto\invr$, agree with data from each regime in each data set, with slightly larger slopes in the FQH regime than in the IQH regime.

\begin{figure}[t!]
\center \label{figure4}
\includegraphics[width=3.25 in]{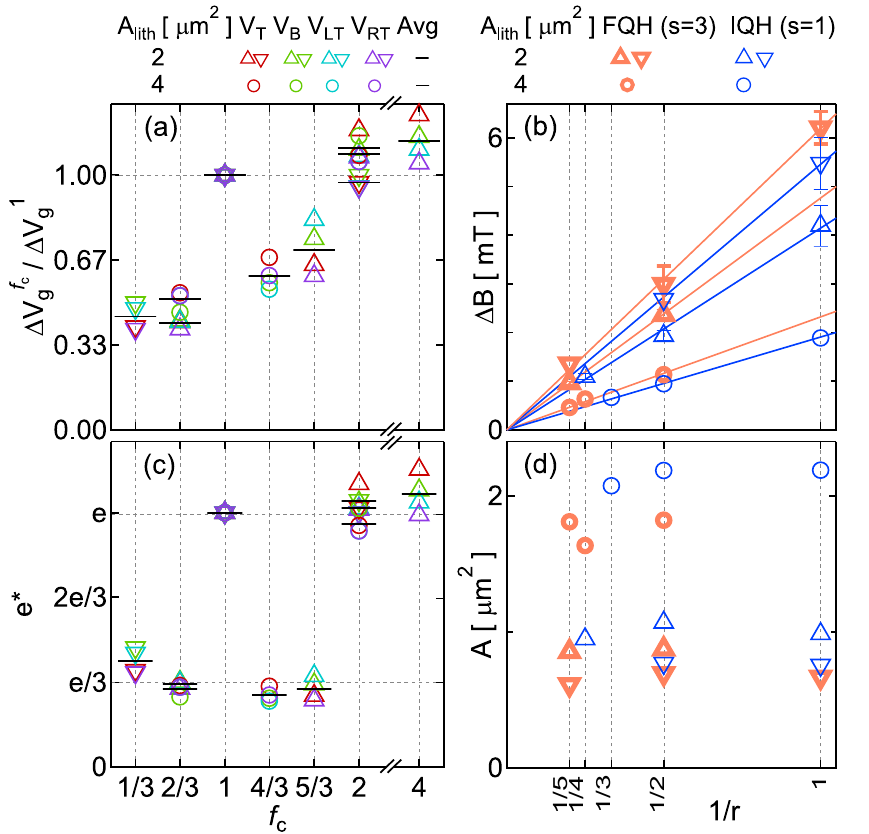}
\caption{\footnotesize{(a) Gate voltage periods $\dvt$ (red), $\dvb$ (green), $\dvlt$ (aqua) and $\dvrt$ (purple), and their average (horizontal black lines), as a function of $\fc$, normalized by their values at $\fc=1$. Here and in all remaining figures, two data sets are taken from the $2~\um^2$ device (triangles) and one from the $4~\um^2$ device (circles), at similar gate voltages for all $\fc$ in each data set. (b) Field periods of IQH ($\fc=r$) [thin, blue] and FQH ($\fc=r/3$) [thick, orange] oscillations versus $\invr$. Error bars, determined by the FWHM of gaussian fits to FFT peaks, are omitted when comparable to marker size or smaller. Fit lines have slopes $1.9, 2.3, 4.1, 4.8, 5.4,$~and~$6.2~\mt$ from bottom to top. (c) Effective charges $\ecs$ extracted from the gate voltage periods shown in (a) using Eq.~1 in the limit of ideal side gates, assuming $\ecs=e$ at $\fc=1$. (d) Effective areas calculated using Eq.~2 from the values of $\Db$ shown in (b).}}
\end{figure}

We next summarize the theoretical charging model~\cite{bernd07,bert10} used to analyze the data. According to this model, resistance oscillations reflect charge-balance considerations of a nearly isolated island of charge, coupled to the leads via forward tunneling, with charging events occurring in units of the quasi-particle charge $\ecs$ in the constrictions~\cite{bernd07, bert10}. When $\fc$ is an integer, $\ecs = e$. When $\fc=r/s$, the composite fermion model~\cite{jainCF} predicts a quasi-particle charge $\ecs=e/s$. The charge on the isolated island is $\nl\ecs$, where $\nl$ is the quantized number of localized quasi-particles. The 2DEG in this area also contains continuous negative charge $\nphi\fc e$ from the underlying IQH or FQH state, where $\nphi=BA/\p0$ is the (non-quantized) number of quanta of flux, $\p0=h/e$, in the area $A$ enclosed by the interfering edge. To minimize energy, the total negative charge must balance the background positive charge $\nbg|e|$ from ionized donors (positive) and gate voltages (negative), yielding the charge neutrality equation $\nphi\fc e+\nl\ecs\approx\nbg e$, where quantization of $\nl$ prevents exact equality. From this result, periods can be obtained by expressing $\nphi$ and $\nbg$ in terms of gate voltage and magnetic field, and finding the change in these parameters needed to induce a unit change in $\nl$.

Gate voltages affect the charge balance in three ways: through the enclosed flux via area, with $\bg\equiv d\nphi/d\vg=(B/\p0)(dA/d\vg)$, and through the background charge via both density $\dbg$ and area. Summing the two background charge effects gives $\gg\equiv d\nbg/d\vg=\dbg(dA/d\vg)+A(d\dbg/d\vg)$, which is assumed $B$-independent~\cite{bert10}. For fixed magnetic field, the charge neutrality equation then yields the gate-voltage period 
\begin{equation}
\dvg=\frac{\ecs/e}{\gg-\bg\fc}.
\end{equation}
This result reflects the Coulomb-blockade intuition that $\dvg\propto\ecs$, but here the gating effect of the underlying Landau levels, represented by $\bg\fc$, may cause the lever-arm to depend on $\fc$: although $\bg\propto B$ and $\fc\sim1/B$, the second relationship is rendered inexact by non-zero plateau widths and the discreteness of $\fc$. In the strong-backscattering limit, therefore, oscillations near weaker plateaus will have larger $\bg\fc$, hence larger $\dvg$, consistent with the data in Fig.~4(a). An $\fc$-independent lever-arm is obtained for $dA/d\vg=0$, i.e.\ for an ideal back gate, but both the geometry of our device and the observed $\fc$-dependence of $\dvg$ suggest that the gates do in fact affect the area.

We simplify the analysis by assuming ideal side gates, which by definition affect area but not density ($d\dbg/d\vg=0$), and an infinitely steep confining potential. In this case $\gg-\bg\fc=\eg(B_1-B\fc)$, where $\eg=\bg/B$ is the only free parameter and $B_1=\nb\p0$ is the field at which $\fb=1$. Then $\eg$ may be extracted from $\dvg$ at a single $\fc$ with known $\ecs$;  we choose $\fc=1$ and use the $\eg$ value extracted there to calculate $\ecs$ at all other $\fc$. Performing this calculation for each gate and each data set yields the values shown in Fig.~4(c), approximately $e/s$ for all $\fc$, with relatively small scatter.

A similar analysis of the charge neutrality equation, assuming fixed gate voltages instead of fixed $B$, predicts a field period
\begin{equation}
\Db=\frac{\p0}{rA},
\end{equation}
where dependence on $\ecs$ has been absorbed by taking $\ecs=e/s$ (justified by the gate-voltage analysis), leaving $A$ as the only fit parameter. As apparent from Fig.~4(d), where Eq.~2 has been used to extract $A$ from each period in Fig.~4(b), FQH-regime periods in each data set indicate slightly smaller areas than corresponding IQH-regime periods, similar to a previous result~\cite{caminoPRL07}. The area difference between the two data sets in the $2~\um^2$ device reflects the use of less-negative gate voltages for the data set with larger areas.

\begin{figure}[t!]
\center \label{figure5}
\includegraphics[width=3.25 in]{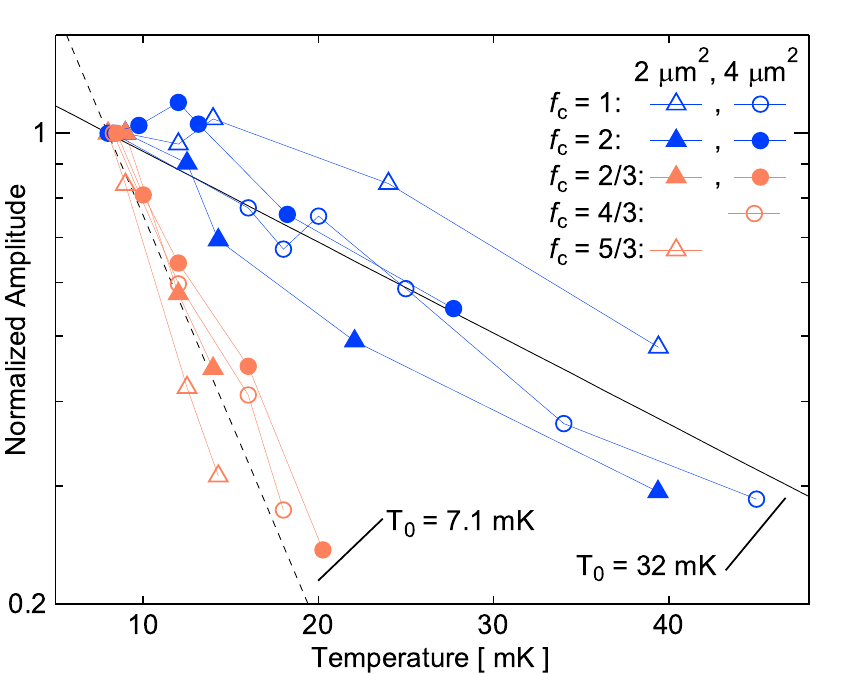}
\caption{\footnotesize{Temperature dependence of oscillation amplitude at several filling factors in the IQH (blue) and FQH (orange) regimes. Lines are given by $D e^{-T / T_0}$, with $T_0$ representing the average value obtained from fits to the individual data sets in each regime. Data at integer $\fc$ have an average $T_0=32~\mk$ and standard deviation $7.0~\mk$, while the fractional $\fc$ have average $T_0=7.1~\mk$ with standard deviation $1.8~\mk$; in both cases, any dependence on filling factor or device size is smaller than the measurement uncertainty. Data at $\fc=4$ were similar to those at $\fc=1,2$ and therefore omitted for clarity, but data at $\fc=1/3$ were unobtainable because of device instability. Above $20~\mk$, FQH-regime oscillations were immeasurably small.}}
\end{figure}

To shed further light on these results, the mixing chamber temperature $T$ is raised and lowered in steps, and oscillations are measured as a function of $B$ after thermal equilibration at each step. The average frequency and amplitude of the oscillations at each $\fc$ were extracted over the same magnetic field range at each temperature. The frequencies were $T$-independent, but the amplitudes depended strongly on $T$, as shown in Fig.~5, where each data set is  normalized by its value at the lowest temperature. Each data set can be characterized by an exponential decay of the form $D e^{-T/T_0}$, where $T_0$ represents a characteristic temperature scale of the oscillations. The continuation of this behavior down to the lowest temperatures confirmed that the 2DEG is well thermalized to the mixing chamber even for $T\lesssim 10~\mk$; furthermore, IQH regime data up to $100~\mk$ (not shown) remained consistent with an exponential dependence, different from the power-law behavior recently observed in the IQH regime at higher temperatures~\cite{hackens10}. The $T_0$ values differed significantly between the IQH and FQH regimes but otherwise appear insensitive to both $\fc$ and area.

A physical interpretation of the exponential dependence and the difference in $T_0$ between the two regimes can be found in Ref.~\cite{bert10}. In this model, $T_0$ reflects an effective charging energy $(\ecs)^2/\ceff$, where $\ceff$ is determined by both the capacitance of the island to ground and edge-structure details. Using this expression with $\ecs=e/s$, the measured $T_0$ yield a value of $\ceff$ twice as large in the IQH regime as in the FQH regime. This difference cannot be attributed directly to the area difference between the two regimes since $T_0$ appears insensitive to area; rather, both differences likely result from a more general structural difference between the IQH and FQH regimes.

In summary, analysis of the gate-voltage periods reveals a quasi-particle charge close to $e/3$ at all FQH states studied, a result that agrees with previous work at $\fc=1/3$, adds to a complicated picture at $\fc=2/3$, and constitutes the first published value at $\fc=4/3$ and $5/3$. The magnetic field periods imply slightly different effective areas for fractional and integer $\fc$. Analysis of the temperature dependence suggests even greater structural differences.

We acknowledge useful discussions with A.\ Kou, B.\ I.\ Halperin, B.\ Rosenow, and I.\ Neder, and funding from Microsoft Corporation Project Q, IBM, NSF (DMR-0501796), and Harvard University. Device fabrication at Harvard Center for Nanoscale Systems.

\small
\bibliographystyle{dougprl}
\bibliography{FQHE}
\end{document}